\documentclass{statsoc}

\usepackage{fullpage}

\usepackage{bm}

\usepackage{graphicx}

\title{Functional modelling of microarray time series with covariate curves}

\author[M. Berk]{Maurice Berk}
\address{Department of Mathematics,
         Imperial College London\\maurice.berk01@imperial.ac.uk}
\email{maurice.berk01@imperial.ac.uk}
\author{Giovanni Montana}
\address{Department of Mathematics,
         Imperial College London\\g.montana@imperial.ac.uk}

\begin{document}

\section{Introduction}
\label{sec:introduction}

Biological systems are inherently dynamic and gene expression levels may be temporally regulated for a wide range of reasons including the cell cycle, circadian rythms, developmental processes or in response to stimuli (e.g. drug treatment or environmental stress) \citep{Spellman1998,Wang2003,Calvano2005}. Microarrays are a high throughput assaying technique for measuring these expression levels of thousands of genes simultaneously. Each microarray hybridisation provides a snapshot of expression levels at a single point in time; by carrying out sequential hybridisations on biological samples arising from the same source (e.g. a human patient), the evolution of these expression levels over time can then be elucidated.

The resulting microarray time series give rise to data that possess certain characteristics which make their analysis particularly challenging. Specifically, due to the large number of genes under study simultaneously, the data is very highly dimensional and there are many more genes than there are time points. Each time series will be replicated typically no more than ten times, and experiments with no replication are not uncommon. The number of genes will often number in the tens of thousands while there are rarely more than ten time points. Even with the falling cost of microarray technology, the limiting factor is often the ability to obtain biological samples which may be restricted due to ethical concerns or other practical, experimental issues. Other challenges include the fact that the data is noisy, with frequent missing observations, and individual heterogeneity.

Our focus is on longitudinal study designs. In this type of microarray experiment, multiple biological units - for example human patients, individual mice or cell lines - are each repeatedly sampled over time to give a collection of observed time series for each gene under study. This type of biological replication is essential for making inference about population parameters but is often overlooked in microarray studies due to experimental issues. A longitudinal microarray experiment is described in Section \ref{sec:data} and provides the data for our case study. The purpose of the study was to follow twelve female and ten male adult human subjects over a period of $6$ months, in order to characterise the change in gene expression levels over time in healthy humans. Figure \ref{fig:raw data} shows some of the raw data for a probe corresponding to the TMEFF1 gene from this example data set where some of the characteristics discussed above can be seen to manifest themselves. A key aspect of human data sets is that the gene expression levels are often collected with covariates - for example, the individual's age, sex and other phenotypic data such as height or weight may be recorded. In the case study, the individuals were stratified by age and gender which allows us to explore not only the evolution of gene expression levels over time but also which genes are differentially expressed between the two gender or age groups. 

When modelling experimental data arising from longitudinal microarray experiments there are three distinct challenges: (a) modelling each individual time series, across all genes and individuals, (b) accounting for the correlation between individuals on a gene by gene basis and (c) modelling the correlation between genes. Accounting for each of these sources of correlation - the temporal, the within-gene (between-individual) and the between-gene - is vital for obtaining better parameter estimates and avoiding a loss of power when testing for genes which are differentially or temporally expressed. With less than 10 timepoints, achieving (a) is not possible with standard time series analysis techniques - it is unlikely, for instance, that we would observe any periodicity. Instead, a field which has proven to be quite successful in this area is that of functional data analysis (FDA). In the FDA paradigm, it is assumed that our observations are noisy realisations of an underlying smooth function of time which is to be estimated. These estimated functions, or curves, are then treated as the fundamental unit of data in any subsequent analysis. Formally, the signal-in-noise model assumed is that observation $y_{i}$ taken at time $t_{i}$ is given by
\begin{eqnarray}
\label{eqn:signal in noise}
& y_{i} = f(t_{i}) + \epsilon_{i} &
\end{eqnarray}
where $f(\cdot)$ is the function of interest to be estimated and $\epsilon_{i}$ is an error term. Typically the infinite dimensional function $f(\cdot)$ is projected onto some finite dimensional basis using parameterisations such as splines, wavelets or fourier bases. In our discussion we will focus on splines in particular as these regularly occur in the literature in terms of both microarray and functional data analysis. For a thorough treatment of FDA, the monograph \cite{Ramsay2005} provides an excellent introduction.

In a longitudinal study, for a particular gene, observations will be collected on not just a single function $f(\cdot)$, but a collection of $n$ functions $f_{i}(\cdot), i=1,\cdots,n$, one for each individual biological unit. Often the main quantity of interest is the population mean function $\mu(\cdot)$ characterising the overall population gene expression level over time. In this case we extend the signal-in-noise model (\ref{eqn:signal in noise}) so that the $j$th observation on individual at time $t_{ij}$ is given by
\begin{eqnarray}
\label{eqn: functional mixed-effects model}
y_{ij} = \mu(t_{ij}) + f_{i}(t_{ij}) + \epsilon_{ij}
\end{eqnarray}
This is known as the functional mixed-effects model and is an extension of the standard linear mixed-effects model \citep{Harville1977} where the fixed- and random-effects are both considered functions. Function $\mu(\cdot)$ is treated as a fixed-effect as it is assumed to be some fixed, but unknown, population function to be estimated. In constrast, the functions $f_{i}(\cdot), i=1,\cdots,n$ represent a random sample from the population as a whole and are assumed to be i.i.d realisations of an underlying stochastic process. Model (\ref{eqn: functional mixed-effects model}) has appeared in a number of different forms depending upon the exact parameterisation of the fixed- and random-effects. For instance, \cite{Guo2002} models both as cubic smoothing splines while \cite{Rice2001} prefer a B-spline representation.

The task of handling correlations amongst the genes has, to date, generally been overlooked by researchers. It is a challenging, open problem to model both the between- and within-gene correlation simultaneously given the size of the data. Although it is well known that genes are co-regulated, for the sake of tractability the most common approach is to simply model each gene independently. In other words, given the framework outlined thus far, each gene would be modelled as a separate functional mixed-effects model.

In this paper we propose a functional-mixed effects model and a framework for estimation and testing in one-sample problems. The model enables the estimation of a mean response curve with the inclusion of covariates, such as gender and sex, also modeled as time-varying smooth functions. We also show how a functional PCA can be applied to the estimated mean curves in order to identify the principal modes of functional variation in the data set, and visually represent the entire set of genes in a low-dimensional plot. 

The structure of the paper is as follow. Section \ref{sec:data} provides a description of a data set, previously collected and analysed by \cite{Karlovich2009}, that we use here as a case study. The proposed model, inferential procedures and functional PCA are provided in Section {\ref{sec:methods}. In Section \ref{sec:results} we present the experimenal results obtained in the context of our case study. In Section \ref{sec:discussion} we discuss how our methodology compares to related models that have appeared in the literature and compare our experimental results to that of the original study, as well as highlight some of the biological implications. Finally, we conclude in Section \ref{sec:conclusions}.

\begin{figure}[htbp]
\includegraphics[width=\textwidth]{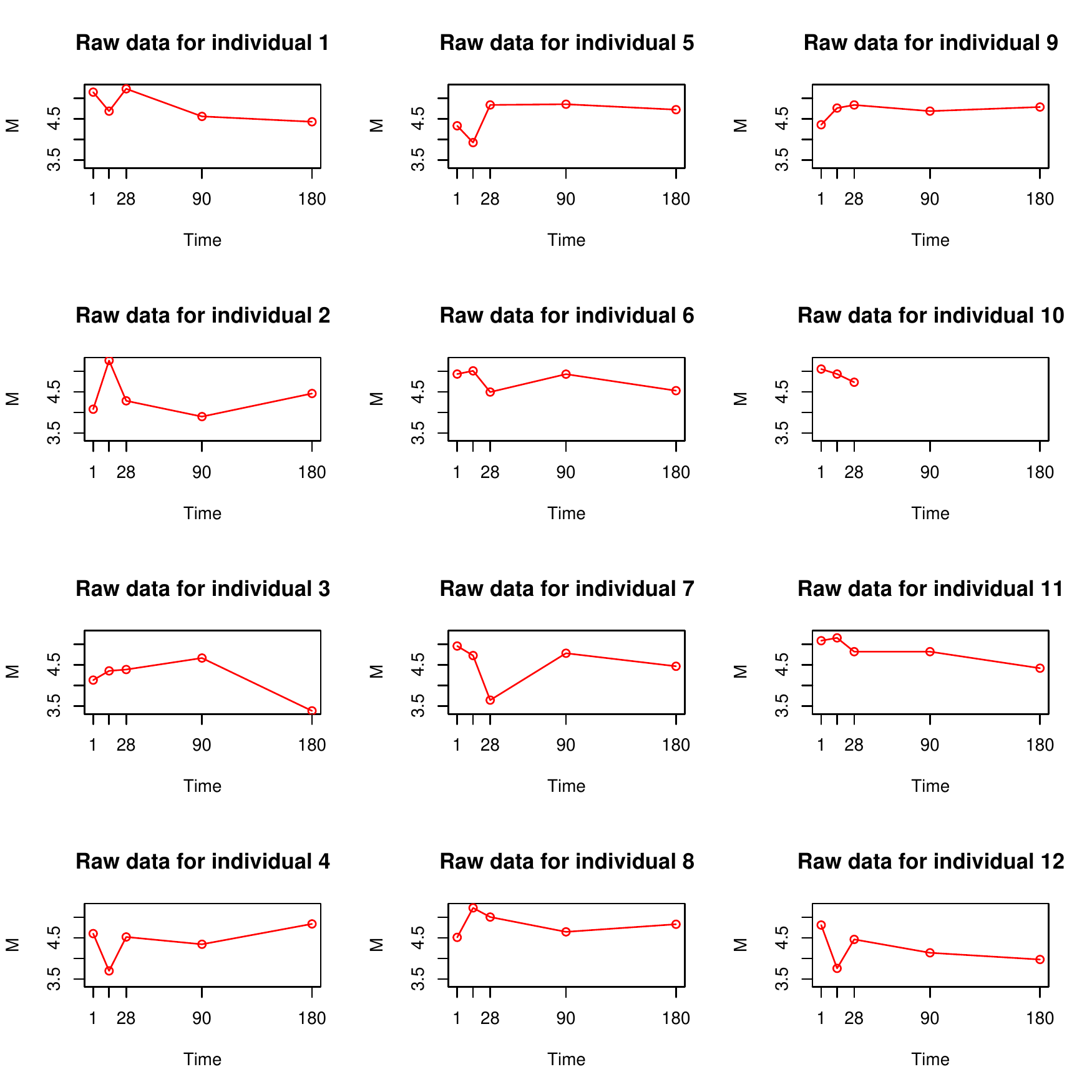}
\caption{\label{fig:raw data}Raw data for TMEFF1, females. Several key characteristics of the data can be observed: (1) irregularly spaced time points (2) missing data - individual 10 is only observed for the first three time points (3) significant individual heterogeneity (4) noisy observations}
\end{figure}

\section{Data description}
\label{sec:data}

The data set used in our case study is taken from \cite{Karlovich2009}. The purpose of the study was to characterise the gene expression levels of healthy human individuals over a period of 6 months. 22 subjects were studied, with gene expression levels assayed from blood samples at days 1, 14, 28, 90 and 180. One subject developed lung cancer during the course of the study and died prior to Day 180, thus contributing only a partial time series. All other individuals completed the study and were observed at all 5 time points. Twelve of the individuals were female and ten were male. In the original study, the subjects were divided into two age groups, with the younger group taken to be those subjects less than or equal to 55 years of age, and the older group those subjects over 55.

In the original paper, the observation for a given gene on individual $i$ at time $t$ was modelled as
\begin{eqnarray*}
y_{it} = \mu + \alpha_{i} + \beta_{g}gender_{i} + \beta_{a}age_{i} + \beta_{t}time_{t} + \epsilon_{it}
\end{eqnarray*}
This is a standard linear mixed-effects model. $\mu$ is the average gene expression level across all individuals after controlling for gender, age and time effects. $\alpha_{i}$ is an individual specific term allowing for a deviation in terms of the intercept of the model. The $\beta_{g}$, $\beta_{a}$ and $\beta_{t}$ parameters separate out the gender, age and time effects respectively while $\epsilon_{it}$ is an error term. 

The model and study design permitted a wide range of biological issue to be explored. Using t-tests, the significance of the age and gender effects was determined. After correcting for multiple-testing by controlling the false discovery rate (FDR) using the procedure of \cite{Benjamini1995}, no genes showed a significant age effect. This was somewhat unexpected given previous studies \citep{Eady2005,Whitney2003,Tang2004} but it was noted that these age effects might be harder to detect in blood than in other tissuses. 78 unique gender genes were identified including XIST, responsible for deactivating one of the X chromosomes in females in order to ensure dosage equivalence, and 23 genes mapped to the Y chromosome. Temporally regulated genes were identified by performing pairwise comparisons between Day 14 and Day 1, Day 28 and Day 14, and Day 180 to Day 90. This was partly due to concerns about a potential batch effect, as Days 1, 14 and 28 were processed in one batch, with Days 90 and 180 being processed in a second batch. No temporally regulated genes were identified in the Day 14 vs Day 1 or the Day 28 vs Day 14 comparisons, but 248 probes were found to be differentially expressed when comparing Day 180 to Day 90, corresponding to 157 unique genes.

Our proposed approach is to replace the original linear mixed-effects model with a functional one. The age and gender effects will be modelled as functions of time, along with the mean and individual curves. To avoid over-parameterisation, all curves will be represented using smoothing splines. The result is a flexible model which permits the interaction of age and gender with time, if the data supports it. During our preprocessing we found little evidence of a batch effect and we will use the entire time course to identify temporally regulated genes, on the basis of the fitted mean function.


\section{Methods}
\label{sec:methods}

We propose the following functional mixed-effects model for the data described in Section \ref{sec:data}. Each gene is modelled independently. For a given gene, the observed gene expression level for individual $i$ at time $t_{ij}$ is given by
\begin{eqnarray}
& y_{i}(t_{ij}) = \mu(t_{ij}) + \alpha_{k}(t_{ij}) + \beta_{l}(t_{ij}) + \gamma_{i}(t_{ij}) + \epsilon_{ij} &
\label{eqn:functional ME}
\end{eqnarray}
where $\mu(\cdot)$ models the mean expression levels across all individuals after accounting for age and gender effects; $\alpha_{k}(\cdot)$ is the gender effect for gender $k$ to which individual $i$ belongs with $k = \{\mbox{Male, Female}\}$; $\beta_{l}(\cdot)$ is the age group effect for group $l$ to which individual $i$ belongs where $l = \{\mbox{Young, Old}\}$; $\gamma_{i}(\cdot)$ is the individual specific effect for individual $i$ and $\epsilon_{ij}$ is an error term. The functions $\mu(\cdot)$, $\alpha_{k}(\cdot)$, $\beta_{l}(\cdot)$ and $\gamma_{i}(\cdot)$ are assumed to be smooth functions of time which we wish to estimate based on the noisy observations. We treat $\mu(\cdot)$, $\alpha_{k}(\cdot)$ and $\beta_{l}(\cdot)$ as fixed-effects, unknown population functions to be estimated, and the $\gamma_{i}(\cdot)$ functions which are treated as random-effects as they represent a random sample of functions from the population as a whole. Formally, the $\gamma_{i}(\cdot)$ are assumed to be i.i.d. realisations of an underlying Gaussian Process with mean 0 and covariance function $\delta(r,s)$.

The functions can be parameterized in a number of ways but we favour smoothing splines as these offer a fine degree of control over the amount to which the data is smoothed. Writing the vector of all observed time points for individual $i$ as $\bm{t}_{i} = [t_{i1}, t_{i2}, \cdots, t_{in_{i}}]^{T}$ where $n_{i}$ is the total number of observations on individual $i$, (\ref{eqn:functional ME}) can be written in matrix form as
\begin{eqnarray}
\label{eqn:functional ME matrix}
& \bm{y}_{i} = \bm{X}_{i}\bm{\mu} + \bm{X}_{i}\bm{\alpha}_{k} + \bm{X}_{i}\bm{\beta}_{l} + \bm{X}_{i}\bm{\gamma}_{i} + \bm{\epsilon}_{i} &
\end{eqnarray}
where $\bm{y}_{i} = [y_{i}(t_{i1}), y_{i}(t_{i2}), \cdots y_{i}(t_{in_{i}})]^{T}$ and $\bm{\epsilon_{i}} = [\epsilon_{i1}, \epsilon_{i2},\cdots,\epsilon_{in_{i}}]^{T}$ are vectors of length $n_{i}$ and $\bm{\mu} = [\mu(\tau_{1}),\mu(\tau_{2}),\cdots,\mu(\tau_{M})]^{T}$ is a vector of length $M$. The vectors $\bm{\alpha}_{k}$, $\bm{\beta}_{l}$ and $\bm{\gamma_{i}}$ are defined similarly to $\bm{\mu}$. The values 
$\tau_{1}, \tau_{2}, \cdots, \tau_{M}$ denote the distinct design time points, of which there are $M$ in total, and $\bm{t}_{i}$ may differ from these may differ if individual $i$ has missing data or duplicate observations for some time points. The matrix $\bm{X}_{i}$ is an incidence matrix of dimension $n_{i} \times M$ where each row $\bm{x}_{ij}$ contains all zeroes aside from the column $m$ where $t_{ij} = \tau_{m}$. Further details on forming the incidence matrices and an example can be found in Appendix \ref{appendix:Xi}. Recall that $\gamma_{i}(\cdot) \sim GP(0,\delta), i=1,\cdots,n$, then the vectors $\bm{\gamma}_{i}$ are multivariate-normally distributed with mean $\bm{0}$ and covariance matrix $\bm{D}$ where $\bm{D}(r,s)=\delta(\tau_{r},\tau_{s})$. Similarly the noise term $\bm{\epsilon}_{i}$ is multivariate-normally distributed with mean $\bm{0}$ and covariance matrix $\bm{R}_{i}$, and we assume that the vectors $\bm{\gamma}_{i}$ and $\bm{\epsilon}_{i}$ are independent.
For simplicty we assume that $\bm{R}_{i} = \sigma^{2}\bm{I}_{n_{i} \times n_{i}}$, although a more complicated structure could be modelled at the expense of fitting more parameters. It is further necessary to impose the identifiability constraint that the age and gender fixed-effects for the two groups sum to zero, i.e. $\bm{\alpha}_{male} + \bm{\alpha}_{female} = \bm{0}$ and $\bm{\beta}_{young} + \bm{\beta}_{old} = \bm{0}$. For simplicity, therefore, we model a single gender and age effect, $\bm{\alpha} = \bm{\alpha}_{female}$ and $\bm{\beta} = \bm{\beta}_{old}$ respectively. These constraints can equivalently be expressed be rewriting (\ref{eqn:functional ME matrix}) as
\begin{eqnarray}
\label{eqn:with constraints}
& \bm{y}_{i} = \bm{X}_{i}\bm{\mu} + \bm{W}_{i}\bm{\alpha} + \bm{Z}_{i}\bm{\beta}_{l} + \bm{X}_{i}\bm{\gamma}_{i} + \bm{\epsilon}_{i} &
\end{eqnarray}
where
\begin{eqnarray*}
& \bm{W}_{i} = \left\{ \begin{array}{ll} -\bm{X}_{i} & \quad \mbox{if \textit{i} is male} \\ \bm{X}_{i} & \quad \mbox{if \textit{i} is female} \end{array} \right. \quad \bm{Z}_{i} = \left\{ \begin{array}{ll} -\bm{X}_{i} & \quad \mbox{if \textit{i} is young} \\ \bm{X}_{i} & \quad \mbox{if \textit{i} is old} \end{array} \right.  & \\
\end{eqnarray*}
Let $\bm{\eta} = [\bm{\mu},\bm{\alpha},\bm{\beta}]^{T}$, then (\ref{eqn:with constraints}) can be rewritten more compactly as
\begin{eqnarray*}
& \bm{y}_{i} = \bm{X}_{i}^{*}\bm{\eta} + \bm{X}_{i}\bm{\gamma}_{i} + \bm{\epsilon}_{i} &
\end{eqnarray*}
where
\begin{eqnarray*}
\nonumber & \bm{X}_{i}^{*} = \left[\begin{array}{ccc}\bm{X}_{i} & \bm{W}_{i} & \bm{Z}_{i}\end{array}\right] &
\end{eqnarray*}
Finally, the complete data vector for all individuals, $\bm{y}$, can be expressed as
\begin{eqnarray}
\label{eqn:lme form}
& \bm{y} = \bm{X}^{*}\bm{\eta} + \widetilde{\bm{X}}\bm{\gamma} + \bm{\epsilon} &
\end{eqnarray}
where $\bm{y} = [\bm{y}_{1}^{T}, \bm{y}_{2}^{T}, \cdots , \bm{y}_{n}^{T}]^{T}$ is an $N = \sum_{i}n_{i}$ length vector, and $\bm{\gamma}$ and $\bm{\epsilon}$ are similarly defined, $\bm{X}^{*} = [{\bm{X}^{*}_{1}}^{T}, {\bm{X}^{*}_{2}}^{T}, \cdots {\bm{X}^{*}_{n}}^{T}]^{T}$ is an $N \times 3M$ matrix  and $\widetilde{\bm{X}} = diag(\bm{X}_{1},\bm{X}_{2},\cdots,\bm{X}_{n})$ is an $N \times nM$ matrix, with the $diag(\cdot)$ operator denoting a block diagonal matrix. The vectors $\bm{\gamma}$ and $\bm{\epsilon}$ are both multivariate-normally distributed with mean $\bm{0}$ and covariance matrix $\widetilde{\bm{D}} = diag(\bm{D},\cdots,\bm{D})$ and $\bm{R} = diag(\bm{R}_{1},\bm{R}_{2},\cdots,\bm{R}_{n})$ respectively.

\subsection{Parameter Estimation}

Model (\ref{eqn:lme form}) is in the form of the standard linear mixed-effects model \citep{Laird1982}. Standard practice for obtaining estimates of the fixed- and random-effects, $\hat{\bm{\eta}}$ and $\hat{\bm{\gamma}_{i}}, i=1,\cdots,n$ would be to maximise the joint likelihood of $\bm{\eta}$ and $\bm{\gamma}_{i}$ \citep{Robinson1991}. This is equivalent to minimising the following generalized log likelihood (GLL) criterion
\begin{eqnarray}
\label{eqn:generalized likelihood criteria}
& GLL = (\bm{y} - \bm{X}^{*}\bm{\eta} - \widetilde{\bm{X}}\bm{\gamma})^{T}\bm{R}^{-1}(\bm{y} - \bm{X}^{*}\bm{\eta} - \widetilde{\bm{X}}\bm{\gamma}) + \log|\widetilde{\bm{D}}| & \\
\nonumber & + \bm{\gamma}^{T}\widetilde{\bm{D}}^{-1}\bm{\gamma} + \log|\bm{R}| &
\end{eqnarray}
However, in our model the fixed- and random-effects are the fitted values of the smoothing spline estimates of the functions $\mu(\cdot)$, $\alpha(\cdot)$, $\beta(\cdot)$, $\gamma_{i}(\cdot), i=1,\cdots,n$, and it is necessary to incorporate a penalty term for the roughness of the smoothing splines into the likelihood. The \textit{penalized} GLL is then given by
\begin{eqnarray}
\label{eqn:penalized likelihood}
& PGLL = GLL + \displaystyle \lambda_{\gamma}\sum_{i=1}^{n}\left\{\int_{a}^{b} [ \gamma_{i}''(t) ]^{2} dt\right\} + \displaystyle \lambda \int_{a}^{b} [ \mu''(t) ]^{2} dt & \\
\nonumber &  + \displaystyle \lambda \int_{a}^{b} [ \alpha''(t) ]^{2} dt + \lambda \int_{a}^{b} [ \beta''(t) ]^{2} dt &
\end{eqnarray}
where the integrals quantify the roughness of the curves $\mu(\cdot)$, $\alpha(\cdot)$, $\beta(\cdot)$, $\gamma_{i}(\cdot), i=1,\cdots,n$ in terms of their squared second derivative, although other penalties could be used. The scalars $\lambda$ and $\lambda_{\gamma}$ are positive-valued smoothing parameters that control the roughness of the fit. For a given smoothing spline fit, $\lambda=0$ would correspond to an interpolation of the data points while as $\lambda$ tends to infinity, the fit tends to a straight line. Note that the same smoothing parameter $\lambda$ is used for the three fixed-effects functions, $\mu(\cdot)$, $\alpha(\cdot)$, $\beta(\cdot)$, and similarly the same smoothing parameter, $\lambda_{\gamma}$, is used for all random-effect functions $\gamma_{i}(\cdot), i=1,\cdots,n$. This is conceptually justified as each function $\gamma_{i}$ is assumed to be a realisation of the same underlying Gaussian Process, but it is possible to envisage selecting a separate smoothing parameter for each fixed- and random-effect function, albeit at the expense of a far greater computational cost.

Minimization of (\ref{eqn:penalized likelihood}) requires calculation of the integral of the squared second derivative of the fixed- and random-effects. In the case of cubic smoothing splines, for a given function $f(t)$ observed at time points $t_{1},t_{2},\cdots,t_{n}$ such that $\bm{f} = [f(t_{1}),f(t_{2}),\cdots,f(t_{n})]^{T}$, there is a \textit{roughness matrix} $\bm{G}$ which can be calculated in a computationally efficient manner that satisfies:
\begin{eqnarray*}
\displaystyle \int_{a}^{b} [f''(t)]^{2}dt = \bm{f}^{T}\bm{G}\bm{f}
\end{eqnarray*}
this result can be found in \cite{Green1994} and we have reproduced the derivation in Appendix \ref{appendix:G} for completeness. Incorporating the roughness matrix into (\ref{eqn:penalized likelihood}) gives
\begin{eqnarray*}
PGLL & = & GLL + \displaystyle \lambda_{\gamma}\sum_{i=1}^{n}\bm{\gamma}_{i}^{T}\bm{G}\bm{\gamma}_{i} + \lambda(\bm{\mu}^{T}\bm{G}\bm{\mu} + \bm{\alpha}^{T}\bm{G}\bm{\alpha} + \bm{\beta}^{T}\bm{G}\bm{\beta}) \\
\nonumber & = & GLL + \displaystyle \lambda_{\gamma}\bm{\gamma}^{T}\widetilde{\bm{G}}\bm{\gamma} + \lambda\bm{\eta}^{T}\bm{G}^{*}\bm{\eta} 
\end{eqnarray*}
where $\widetilde{\bm{G}}$ is a block diagonal matrix comprised of the matrix $\bm{G}$ repeated $n$ times. Similarly, $\bm{G}^{*}$ is a block diagonal matrix comprised of $\bm{G}$ repeated three times.

After a rearrangement on the terms featuring in the penalised log-likelihood, the model can be re-written in terms of the \textit{regularised} covariance matrices $\widetilde{\bm{D}}_{\gamma} = (\widetilde{\bm{D}}^{-1} + \lambda_{\gamma}\widetilde{\bm{G}})^{-1}$ and $\bm{V} = \widetilde{\bm{X}}\widetilde{\bm{D}}_{\gamma}\widetilde{\bm{X}}^{T} + \bm{R}$, so called because the matrix $\widetilde{\bm{D}}_{\gamma}$ is obtained by regularising the covariance matrix $\widetilde{\bm{D}}$ with the term $\lambda_{\gamma}\widetilde{\bm{G}}$. This method of imposing the smoothness constraints by regularisation of the covariance matrix can be credited to \cite{Wu2006}.




Minimising (\ref{eqn:penalized likelihood}) gives the BLUE and BLUP of the fixed- and random-effects as
\begin{eqnarray}
\label{eqn:BLUE and BLUP}
& \hat{\bm{\eta}} = ({\bm{X}^{*}}^{T}\bm{V}^{-1}\bm{X}^{*} + \lambda\bm{G}^{*})^{-1}{\bm{X}^{*}}^{T}\bm{V}^{-1}\bm{y} & \\
& \hat{\bm{\gamma}} = \widetilde{\bm{D}}_{\gamma}\widetilde{\bm{X}}^{T}\bm{V}^{-1}(y - \bm{X}^{*}\bm{\eta}) &
\end{eqnarray}
The discussion thus far has assumed that the variance components $\bm{D}$ and $\sigma^{2}$ were known. Of course, in practical applications this will not be the case. Assuming the random-effects $\bm{\gamma}_{i}$ and error terms $\bm{\epsilon}$ are known, the maximum likelihood estimators $\hat{\bm{D}}$ and $\hat{\sigma}^{2}$ are given as
\begin{eqnarray}
\label{eqn: MLE variance components}
& \hat{\bm{D}} = \displaystyle \frac{1}{n}\sum_{i=1}^{n}\bm{\gamma}_{i}\bm{\gamma}_{i}^{T} \quad \hat{\sigma}^{2} = \frac{1}{N} \bm{\epsilon}^{T}\bm{\epsilon}&
\end{eqnarray}
As the random-effects $\bm{\gamma}_{i}$ and error terms are not, in fact, directly observed, we resort to the Expectation-Maximisation algorithm where they can be treated as missing data. In this procedure the sufficient statistics of $\hat{\bm{D}}$ and $\hat{\sigma}^{2}$ - $\bm{\gamma}_{i}\bm{\gamma}_{i}^{T}, i=1,\cdots,n$ and $\bm{\epsilon}^{T}\bm{\epsilon}$ respectively - are replaced by their conditional expectations which are calculated at the E-step. In the M-step, the maximum likelihood estimators are then calculated having replaced the sufficient statistics by these conditional expectations, which are given by
\begin{eqnarray}
\label{eqn: conditional expectation1}
E[\bm{\gamma}_{i}\bm{\gamma}_{i}^{T}|\bm{y},\bm{\eta}=\hat{\bm{\eta}}] & = & \hat{\bm{\gamma}}_{i}\hat{\bm{\gamma}}_{i}^{T} + \hat{\bm{D}}_{\gamma} - \hat{\bm{D}}_{\gamma}\bm{X}_{i}^{T}\bm{V}_{i}^{-1}\bm{X}_{i}\hat{\bm{D}}_{\gamma}\\
\label{eqn: conditional expectation2}
E[\bm{\epsilon}^{T}\bm{\epsilon}|\bm{y},\bm{\eta}=\hat{\bm{\eta}}] & = & \hat{\bm{\epsilon}}^{T}\hat{\bm{\epsilon}} + \hat{\sigma}^{2}N - \hat{\sigma}^{4}tr(\bm{V}^{-1})
\end{eqnarray}
where $tr(\cdot)$ denotes the trace of a matrix and $\bm{V}_{i} = \bm{X}_{i}\bm{D}_{\gamma}\bm{X}_{i}^{T} + \sigma^{2}\bm{I}_{n_{i} \times n_{i}}$. Derivations of these conditional expectations are given in Appendix \ref{appendix:conditional expectations}.


\subsection{Model Selection}

Thus far we have treated the smoothing parameters $\lambda$ and $\lambda_{\gamma}$ as fixed. In reality, optimal values of these parameters must be found using a model selection procedure. \cite{Guo2002} made use of the relationship between a smoothing spline and a linear mixed-effects model in order to treat the smoothing parameters as variances components that could be estimated during the normal course of the EM-algorithm. We prefer, however, to dissociate the model selection from parameter estimation and numerically optimise over the two dimensional space of non-negative reals ($\Lambda \times \Lambda_{\gamma}$) as this is a much more flexible approach. There are a number of different criteria for scoring the smoothing parameters, all of which essentially trade off between model fit and model complexity.

\cite{Ma2006}'s smoothing-spline clustering approach for microarray data, for instance, employed \cite{Wahba1977}'s generalized cross validation (GCV) criterion. It is well known, however, that GCV tends to undersmooth \citep{Lee2003}. Alternatively, we can employ either the Akaike Information Criterion (AIC) or the Bayesian Information Criterion (BIC):
\begin{eqnarray*}
& AIC(\lambda,\lambda_{\gamma}) = -2 \mbox{lik} + 2 \mbox{df} &\\
& BIC(\lambda,\lambda_{\gamma}) = -2 \mbox{lik} + log(N) \mbox{df} &
\end{eqnarray*}
These two criteria both score the smoothing parameters in terms of the likelihood - measuring the model fit - adjusted for a penalty term for the model complexity, in terms of degrees of freedom. The difference lies in the size of the penalty term, with BIC giving more conservative results when $log(N) > 2$, in other words when there are more than 9 data points. 


Both of these criteria, and GCV, have a sound theoretical basis. We suggest, therefore, to choose which one to use on the basis of \textit{a priori} knowledge about the kind of patterns we expect to observe in a given data set. If, as in our example data set, we do not expect there to be many genes with curvy temporal profiles, then we may prefer the more conservative BIC. On the other hand, in a data set with a greater number of time points and with more expected variability - in response to infection for instance - then we may prefer the AIC in order to better capture the more complex patterns expected.

\subsubsection{Smoother Matrices}

In order to evaluate the criteria, it is necessary to calculate the degrees of freedom of the model. As per \cite{Buja1989}, the degrees of freedom associated with the fixed- and random-effects, $\bm{\eta}$ and $\bm{\gamma}$, can be expressed as the trace of some smoother matrix $\bm{A}$ such that $\hat{\bm{y}} = \bm{A}\bm{y}$. Equivalently, it is useful to determine the two smoother matrices $\bm{A} = \bm{A}_{\eta} + \bm{A}_{\gamma}$ so that the degrees of freedom of the fixed- and random-effects can be accounted for separately.

Recall that the fitted values of the fixed-effects at the design time points can be written as $\bm{X}^{*}\hat{\bm{\eta}}$. Replacing $\hat{\bm{\eta}}$ with (\ref{eqn:BLUE and BLUP}) gives
\begin{eqnarray*}
& \bm{X}^{*}\hat{\bm{\eta}} = \bm{X}^{*}({\bm{X}^{*}}^{T}\bm{V}^{-1}\bm{X}^{*} + \lambda\bm{G}^{*})^{-1}{\bm{X}^{*}}^{T}\bm{V}^{-1}\bm{y} = \bm{A}_{\eta}\bm{y} &
\end{eqnarray*}
and so the smoother matrix $\bm{A}_{\eta}$ is given by
\begin{eqnarray*}
& \bm{A}_{\eta} = \bm{X}^{*}({\bm{X}^{*}}^{T}\bm{V}^{-1}\bm{X}^{*} + \lambda\bm{G}^{*})^{-1}{\bm{X}^{*}}^{T}\bm{V}^{-1} &
\end{eqnarray*}
Similarly, the fitted values of the random-effects at the design time points can be written as $\widetilde{\bm{X}}\hat{\bm{\gamma}}$, which gives
\begin{eqnarray*}
& \widetilde{\bm{X}}\hat{\bm{\gamma}} = \widetilde{\bm{X}}\widetilde{\bm{D}}_{\gamma}\widetilde{\bm{X}}^{T}\bm{V}^{-1}(\bm{I}_{N} - \bm{A}_{\eta})\bm{y} = \bm{A}_{\gamma}\bm{y} &
\end{eqnarray*}
The degrees of freedom of the model can then be calculated as $df = tr(\bm{A}_{\eta} + \bm{A}_{\gamma}) + 1$, which is the trace of the smoother matrix plus an additional paramter for fitting the noise variance $\sigma^{2}$.

With the scoring function in place any kind of two-dimensional optimisation routine can be used, although in practice a simple grid search or sequential line optimisation is recommended \citep{Wu2006}. We have found that a more sophisticated simplex-search optimiser \citep{Nelder1965} can be employed without incurring a significant computational cost. This allows optimisation over the two smoothing parameters $\lambda$ and $\lambda_{\gamma}$ simultaneously without needing to calculate the derivative of the criterion.

\subsection{Confidence Bands}

Pointwise confidence bands at the design time points for each of the fixed-effects functions can be determined either theoretically or using a bootstrap resampling procedure. In the case of the former, we have
\begin{eqnarray*}
& cov(\hat{\bm{\eta}}) = ({\bm{X}^{*}}^{T}\bm{V}^{-1}\bm{X}^{*} + \lambda\bm{G}^{*})^{-1}{\bm{X}^{*}}^{T}\bm{V}^{-1}\bm{X}^{*}({\bm{X}^{*}}^{T}\bm{V}^{-1}\bm{X}^{*} + \lambda\bm{G}^{*})^{-1} &
\end{eqnarray*}
The diagonal elements of $cov(\hat{\bm{\eta}})$, therefore, give the variance of the fixed-effects at the design time points with the first $M$ elements corresponding to $\mu(\cdot)$, the next $M$ elements to $\alpha(\cdot)$, and the final $M$ elements to $\beta(\cdot)$. In fact, due to the block diagonal structure of $cov(\hat{\bm{\eta}})$, these $M$ elements will be the same across all three fixed-effects. Confidence bands for a significance level $\alpha$ at the design time points $\tau_{i}$ can then be calculated for $\hat{\bm{\mu}}$ as $\hat{\mu}(\tau_{i})\pm z \sqrt{cov(\hat{\mu}(\tau_{i}))}$, where $z$ is the critical value under the normality assumption such that $\phi(z) = 1 - \frac{1}{2}\alpha$. These bands can be calculated for the other fixed-effects $\hat{\bm{\alpha}}$ and $\hat{\bm{\beta}}$ in an identical fashion.

Alternatively, confidence intervals can be estimated by resampling the between- and within-individual residuals. To construct a bootstrapped sample for a single individual, first one of the individual functions $\gamma_{i}$ is randomly selected and evaluated at the design time points - denote this vector as $\bm{\gamma}^{*}$. Next, $M$ residuals from the noise vector $\bm{\epsilon}$, are resampled with replacement, writing this vector as $\bm{\epsilon}^{*}$. Then, the bootstrapped observation vector $\bm{y}^{*}$ is given by
\begin{eqnarray*}
& \bm{y}^{*} = \bm{\mu} + \bm{\alpha}^{*} + \bm{\beta}^{*} + \bm{\gamma}^{*} + \bm{\epsilon}^{*} &
\end{eqnarray*}
where $\bm{\alpha}^{*}=\bm{\alpha}$ if the individual is female and $-\bm{\alpha}$ otherwise, similarly for $\bm{\beta}$. This process is then repeated for $n$ individuals, sampling the individual functions with replacement, to give a complete bootstrapped data set. The model is then fit to this resampled data and new estimates for the fixed-effects obtained. Repeating this process for a large number of iterations gives a large number of fixed-effects estimates from which the confidence bands at a given significance level can be determined empirically.

\subsection{Hypothesis Testing}

Fitting model (\ref{eqn:lme form}) allows us to separate out the mean, age and gender effects for each gene. It is then possible to determine whether there is a significant difference between age groups or genders for a given gene by testing the null hypothesis that the size of the effect is zero. As the effects are modelled as functions, a natural way to quantify their size is the  $L_{2}$ norm and testing the significance of the age effect for a given gene can be framed as
$$
H_{0}: ||\alpha(\cdot)||_{2} = 0 \qquad
H_{1}: ||\alpha(\cdot)||_{2} > 0 
$$
Assessing the significance is complicated by the fact that the sample sizes are small and the null distribution is unknown. Instead, it can be determined empirically by using a data resampling scheme such as the bootstrap or permutation procedure. For example, in \cite{Storey2005}, the null distribution of their F-type test-statistic was determined using a nonparametric bootstrap procedure by resampling the individual effects and error terms with replacement. Here, however, the null distribution of the $L_{2}$ norm of the age and gender effects can be estimated empirically using a permutation procedure where the class assignments - male/female or young/old - are randomly permuted.

These same ideas can be applied when testing for genes which are temporally regulated. In this case, the null hypothesis that there is no change over time can be formulated as $||\mu'(\cdot)||_{2} = 0$ where $\mu'(\cdot)$ is the first derivative of the mean curve. The null distribution can then be empirically estimated by randomly permuting the time points.

\subsection{Functional Principal Components Analysis}

Fitting model (\ref{eqn:lme form}) to each gene yields a set of mean curves $\mu_{i}(t), i=1,\cdots,G$ where $G$ is the total number of genes in the data set. Performing a fPCA on this set of curves allows us to identify the main patterns of variation across all genes. A straight-forward way of doing so is the discretisation method described in \cite{Ramsay2005} which is essentially a two-stage approach to fPCA: (1) the data are smoothed by fitting model (\ref{eqn:lme form}) to each gene (2) a fPCA is performed on the smoothed data - in the form of the set of curves $\mu_{i}(t), i=1,\cdots,G$. Alternative methods of fPCA such as \cite{James2000}, which estimate and smooth the PCs directly, cannot be applied in this case where there are two levels of variation - the between and within-gene.

First, each curve is discretised on a fine grid of $n$ equally spaced points across the range of the time course. If there are $N$ curves in total, this yields a data matrix $\bm{X}$, of dimension $N \times n$. A standard PCA can then be performed on $\bm{X}$.

This procedure gives $n$ principal components, each a vector of length $n$. It is then necessary to transform these vectors back into functions. A standard PCA can be defined as solving the eigenequation
\begin{eqnarray}
\label{eqn:standard PCA}
& \bm{V}\bm{u} = \lambda\bm{u} &
\end{eqnarray}
where $\bm{V} = N^{-1}\bm{X}^{T}\bm{X}$ is the sample covariance matrix of $\bm{X}$, $\lambda$ is one of the eigenvalues of $\bm{V}$, and $\bm{u}$ is one of the eigenvectors, or principal components. In the functional setting, we replace $\bm{V}$ by a covariance function $v(s,t)$, and $\bm{u}$ by a function of $s$, $\xi(s)$ such that the eigenequation (\ref{eqn:standard PCA}) becomes
\begin{eqnarray}
\label{eqn:functional PCA}
& \displaystyle \int v(s,t)\xi(t) dt = \rho \xi(s) &
\end{eqnarray}
for a given value of $s$. Noting that after discretisation of the curves the elements of the matrix $\bm{V} = v(s_{j},s_{k})$ where $j$ and $k$ are any of the $n$ discretised points on the fine grid, the integral in (\ref{eqn:functional PCA}) can be approximated as a summation such that
\begin{eqnarray*}
& \displaystyle \int v(s,t)\xi(t) dt = w \sum_{k=1}^{n} v(s,s_{k}) \tilde{\xi}_{k} &
\end{eqnarray*}
where $w$ is the spacing between the points on the fine grid, and $\tilde{\xi}_{k}$ are the discretised values of the function $\xi(s)$. The approximate discrete form of the functional eigenequation is therefore
\begin{eqnarray*}
& w \bm{V} \tilde{\bm{\xi}} = \rho \tilde{\bm{\xi}}&
\end{eqnarray*}
which corresponds to (\ref{eqn:standard PCA}) with $\rho = w\lambda$. Assuming the eigenvectors obtained from the standard PCA have been normalised, the equivalent functional constraint that $\int \xi(s)^{2}ds = 1$ is achieved by enforicing $w||\tilde{\bm{\xi}}||^{2}=1$. The function $\xi(\cdot)$ is then recovered by interpolating the points $\tilde{\bm{\xi}}$. Assuming the grid is fine enough, the choice of interpolation method is irrelevant.

As with a standard PCA, we will wish to retain only a small number of functional PCs. As is standard practice, the eigenvalues $\rho$ can be used to facilitate this choice, by retaining enough PCs to explain most of the variation in the data. Assuming $K$ PCs are retained, for curve $i$ we have
\begin{eqnarray*}
& y_{i}(t) = \displaystyle \mu(t) + \sum_{k}^{K} \kappa_{ik}\hat{\xi}_{k}(t) + \epsilon_{i}(t) &
\end{eqnarray*}
where $\kappa_{ik}$ are the PC loadings for curve $i$. These can be estimated by minimising the residuals $y_{i}(t) - \sum_{k}^{K} \kappa_{ik}\hat{\xi}_{k}(t)$, which in practice again requires discretisation of the curve $i$, and the PCs $\hat{\xi}_{k}(t)$.

\section{Results}
\label{sec:results}

We fit the functional mixed-effects model described in Section \ref{sec:methods} to the example data set described in Section \ref{sec:data}, independently for each probe. Convergence of the EM algorithm was confirmed by convergence of the variance components estimates $\hat{\sigma}^{2}$ and $\hat{\bm{D}}$ and typically took around 30 iterations. 100 iterations of the simplex optimisation procedure were used to select the smoothing parameters. After obtaining estimates of the mean, age and gender effects, and individual curves, these were assessed for significance. To relieve some of the computational burden, permuted null test statistics were shared across all genes - theoretical results justifying this pooling can be found in \cite{Storey2004}. Each gene was permuted 32 times, yielding in excess of 1 million null test statistics for each comparison. From these null distributions, empirical p-values were calculated, which were then corrected for multiple testing using the procedure of \cite{Benjamini1995} to control the FDR at 10\%.

After applying multiple testing corrections, no significant age genes were identified, as in the original analysis. 21 probes were found to be gender specific. Two of these 21 probes can be found on the Y-chromosome but are not mapped to any known genes. The remaining probes correspond to 7 known genes and 2 open reading frames, given in Table \ref{table:gender genes}. Aside from XIST which, as discussed in Section \ref{sec:data} is only expressed in females and is responsible for X-chromose inactivation to facilitate dosage equivalence between the sexes, all significant genes and the two open reading frames are found on the Y-chromosome.

The highest ranked gender-effect gene on an autosomal chromosome was found to be TUBB2A, located on chromosome 6 and ranked number 23, with an associated FDR of 13\%, hence of borderline significance. The gene and fitted mean and gender-effect curves is plotted in Figure \ref{fig:tubb2a}, where a definite difference between the two groups is apparent, corresponding to between a 3- and 4-fold difference in expression levels.

299 probes were found to be significantly temporally regulated, corresponding to 183 unique, mapped genes. The highest ranking gene was found to be MBP - myelin basic protein - given as one of the examples in Figure \ref{fig:outliers}. Myelin is an insulating sheath covering nerve cells, essential for the correct functioning of the central nervous system and degredation of myelin can be found in many neurodegenerative diseases such as multiple sclerosis. It is thought that MBP might function to maintain the correct structure of myelin, which may explain why we found it to be seasonally regulated, although we could find no existing evidence of this.

We performed a functional PCA of the gene mean curves. Each curve was discretised into 1,000 equally spaced points, then normalised by subtracting the first observation from the rest of the points. Thus, each curve represents the change in expression levels over time, relative to $t=0$. The first two PC functions are given in Figure \ref{fig:functional PCs}. The first PC accounts for 99.4\% of the variation and corresponds to a linear change in expression levels over time. The second PC accounts for 0.5\% of the variation and describes expression levels which rise over the first threee months before falling for the next three months, or vice versa. As these two PCs represent almost all of the variation in the curves, we estimated the loadings for each gene and plotted the results in Figure \ref{fig:loadings}. Four outliers have been highlighted and each of these is plotted in Figure \ref{fig:outliers}. It can be seen that the outliers in the loadings plot correspond to those genes which change most over time, with the distinctive line of points in the center corresponding to genes which change linearly. For these genes with linear dynamics, the size of the first PC loading is relative to the slope. Genes which can be separated on the y-axis are those with a quadratic temporal profile.

\begin{table}
\caption{\label{table:karlovich gender genes}19 probes found to be significantly differentially expressed according to gender by \cite{Karlovich2009}, with a mean log-transformed signal intensity greater than or equal to 7}
\centering
\small
\fbox{%
\begin{tabular}{|l|l|l|l|}
\hline
Gene Name & Chromosome & Affymetrix ID & Fold Change\\
\hline
\hline
- & - & 211074\_at & 0.82\\
EIF1AX & X & 201019\_s\_at & 0.86\\
TMEFF2 & M & 224321\_at & 0.87\\
FLOT1 & 6 & 210142\_x\_at & 0.87\\
EIF2S3 & X & 224936\_at & 0.90\\
RPS4X & X & 213347\_x\_at & 0.91\\
MGC71993 & 17 & 224573\_at & 0.93\\
EEF1A1 & 1 & 213477\_x\_at & 1.05\\
EEF1A1 & 6 & 206559\_x\_at & 1.07\\
SPOP & 17 & 204640\_s\_at & 1.07\\
ERBB2IP & 5 & 217941\_s\_at & 1.09\\
UHMK1 & 1 & 224691\_at & 1.11\\
PP784 & 4 & 212199\_at & 1.12\\
HMGN4 & 6 & 209787\_s\_at & 1.13\\
C10orf45 & 10 & 223058\_at & 1.13\\
HTATSF1 & X & 202602\_s\_at & 1.14\\
GNG2 & 14 & 224964\_s\_at & 1.14\\
HMGN4 & 6 & 209786\_at & 1.17\\
HMGN4 & 6 & 202579\_x\_at & 1.20\\
\hline
\end{tabular}}
\end{table}

\begin{table}
\caption{\label{table:gender genes}21 probes found to have a significant gender-effect Aside from XIST, all of these probes can be found on the Y-chromosome. Q-value indicates the corresponding false discovery rate (FDR) if a particular gene is taken to be the cut-off between significant and non-significant.}
\centering
\small
\fbox{%
\begin{tabular}{|l|l|l|p{1.5cm}|l|}
\hline
Gene Name & Chromosome & Affymetrix ID & $L_{2}$ norm & q-value\\
\hline
\hline
XIST & X & 224588\_at & 57.2 & 0.00248\\
XIST & X & 224590\_at & 53.9 & 0.00248\\
EIF1AY & Y & 204409\_s\_at & 48.8 & 0.00248\\
RPS4Y1 & Y & 201909\_at & 42.9 & 0.00248\\
DDX3Y & Y & 205000\_at & 36.7 & 0.00248\\
XIST & X & 214218\_s\_at & 35.4 & 0.00248\\
EIF1AY & Y & 204410\_at & 34.5 & 0.00248\\
XIST & X & 221728\_x\_at & 33.2 & 0.00248\\
CYorf15B & Y & 214131\_at & 30.3 & 0.00248\\
CYorf15A & Y & 232618\_at & 29.2 & 0.00248\\
USP9Y & Y & 228492\_at & 27.8 & 0.00248\\
JARID1D & Y & 206700\_s\_at & 25.3 & 0.00248\\
XIST & X & 224589\_at & 24.8 & 0.00248\\
- & Y & 244482\_at & 22.4 & 0.00430\\
XIST & X & 227671\_at & 22.2 & 0.00430\\
TSIX & X & 231592\_at & 18.4 & 0.0247\\
BCORL2 & Y & 1562313\_at & 18.4 & 0.0247\\
- & Y & 1560800\_at & 16.3 & 0.0323\\
DDX3Y & Y & 205001\_s\_at & 16.1 & 0.0543\\
CYorf15B & Y & 223646\_s\_at & 14.1 & 0.0597\\
CYorf15A & Y & 236694\_at & 13.8 & 0.0845\\
\hline
\end{tabular}}
\end{table}

\begin{figure}[htbp]
\includegraphics[width=\textwidth]{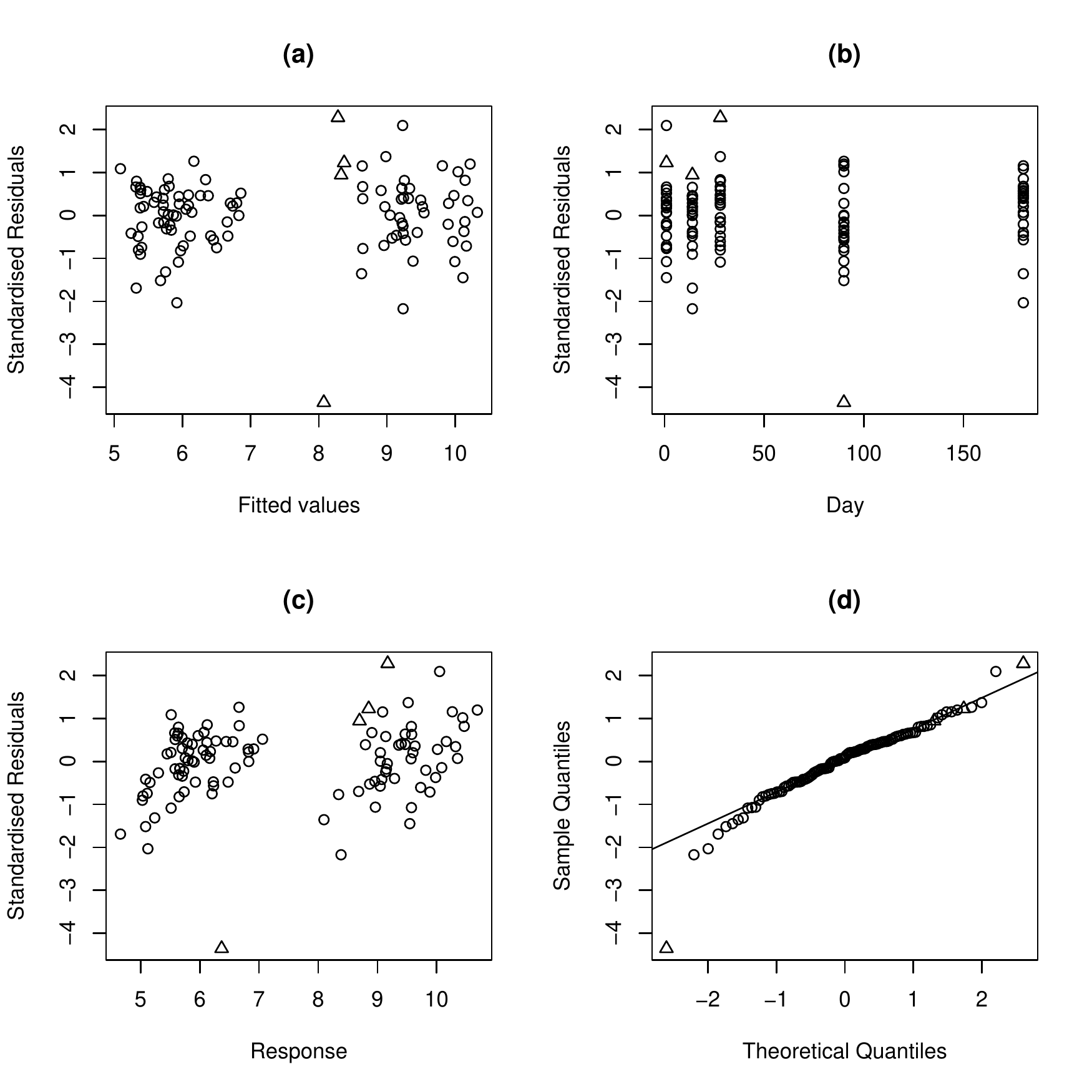}
\caption{\label{fig:residual analysis}Residual analysis for the TUBB2A model fit. (a) Standardised residuals against fitted values (b) Standardised residuals against time (c) Standardised residuals against observations (d) QQ-plot of standardised residuals. These plots can be used to detect patterns in the data which the model has failed to capture. Aside from the obvious groupings as a result of the difference in gene expression levels between males and females, there appears to be little structure to the residuals. In all cases, the triangles correspond to observations on subject 174, who developed lung cancer during the course of the study and died prior to the final time point. It can be seen that this subject contributes two obvious outlying residuals, which may have negatively impacted the goodness of fit criteria calculated by \cite{Karlovich2009}, possibly resulting in its removal from any subsequent analysis}
\end{figure}

\begin{figure}[htbp]
\includegraphics[width=\textwidth]{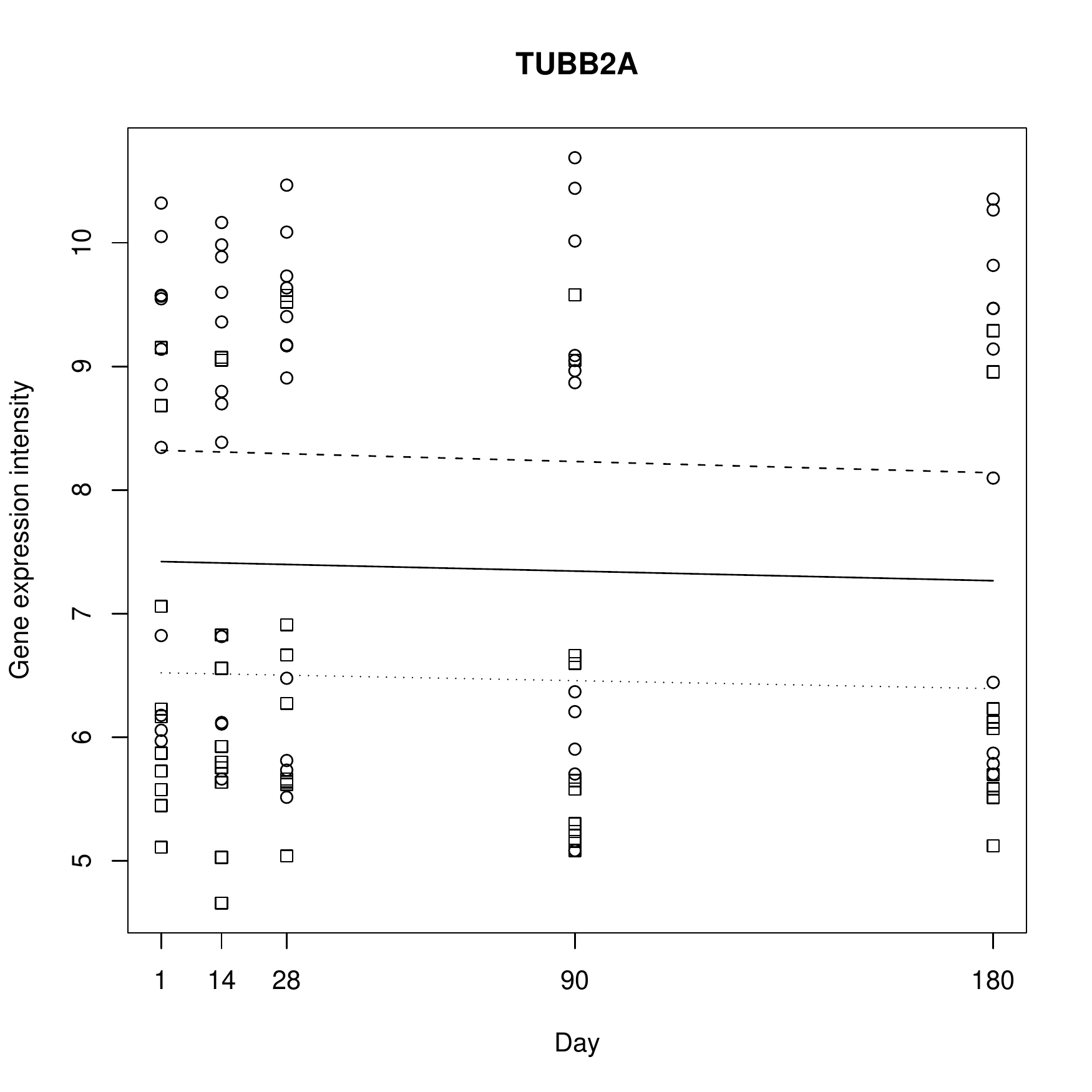}
\caption{\label{fig:tubb2a}Plot of TUBB2A's fitted longitudinal profiles. We have identified TUBB2A as a gene with a potentially novel gender effect. Observations on females are shown as squares, and those on males are shown as circles. The solid line is the overall mean expression level over time, after removing age and gender effects. The dotted line is the mean plus gender effect for females, and the dashed line is the mean plus gender effect for males}
\end{figure}

\begin{figure}[htbp]
\includegraphics[width=\textwidth]{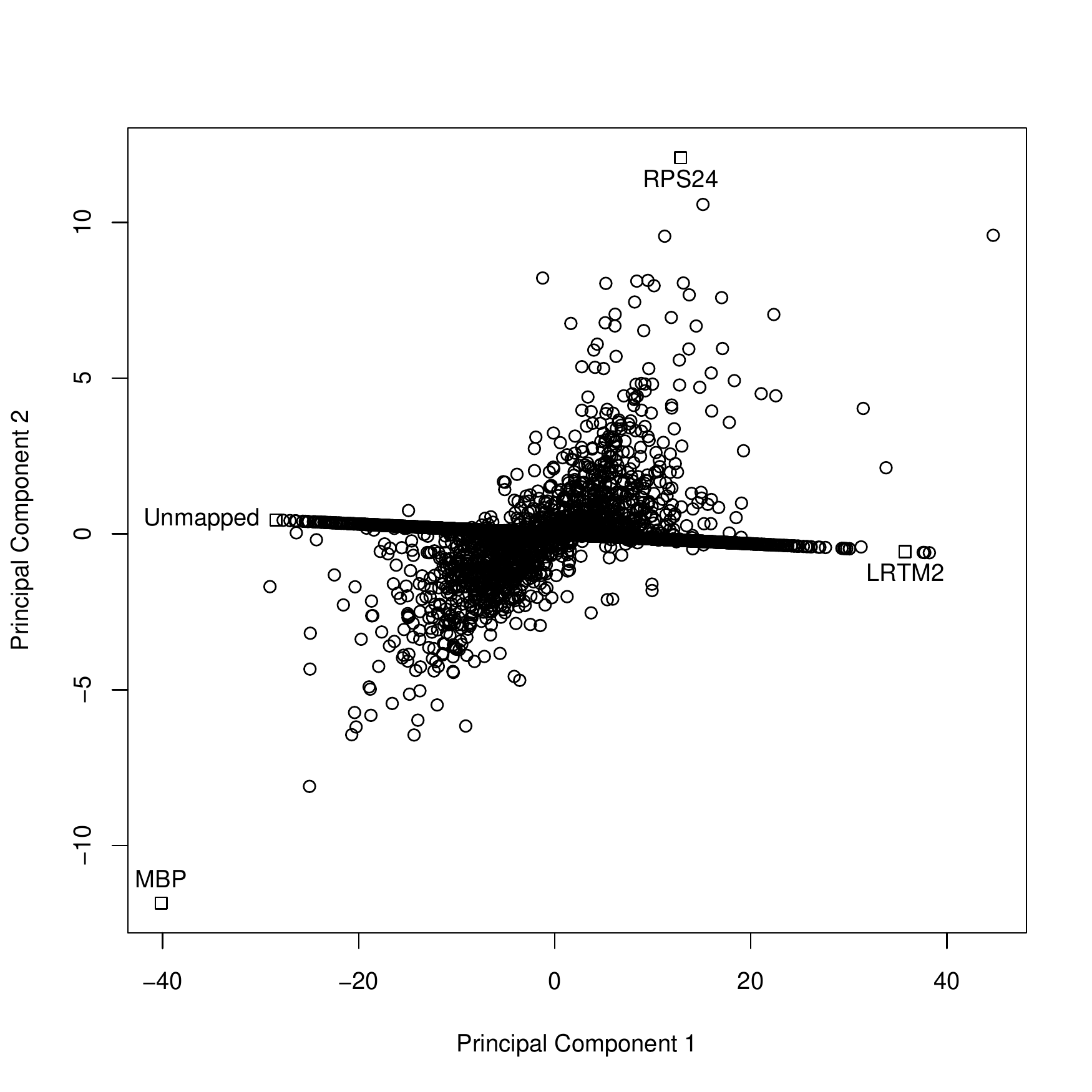}
\caption{\label{fig:loadings}Functional principal components analysis loadings plot. Two functional principal components capture 99.9\% of the observed variation in the fitted mean curves for each gene. The loadings on the first principal component function corresponds to the x-axis, which represents linear variation over time. The second principal component function captures variation which is of a more quadratic nature. These two principal component functions are given in Figure \ref{fig:functional PCs}. Four outliers representing the spectrum of observed temporal profiles have been highlighted; individual plots for these genes are given in Figure \ref{fig:outliers}}
\end{figure}

\begin{figure}[htbp]
\includegraphics[width=\textwidth]{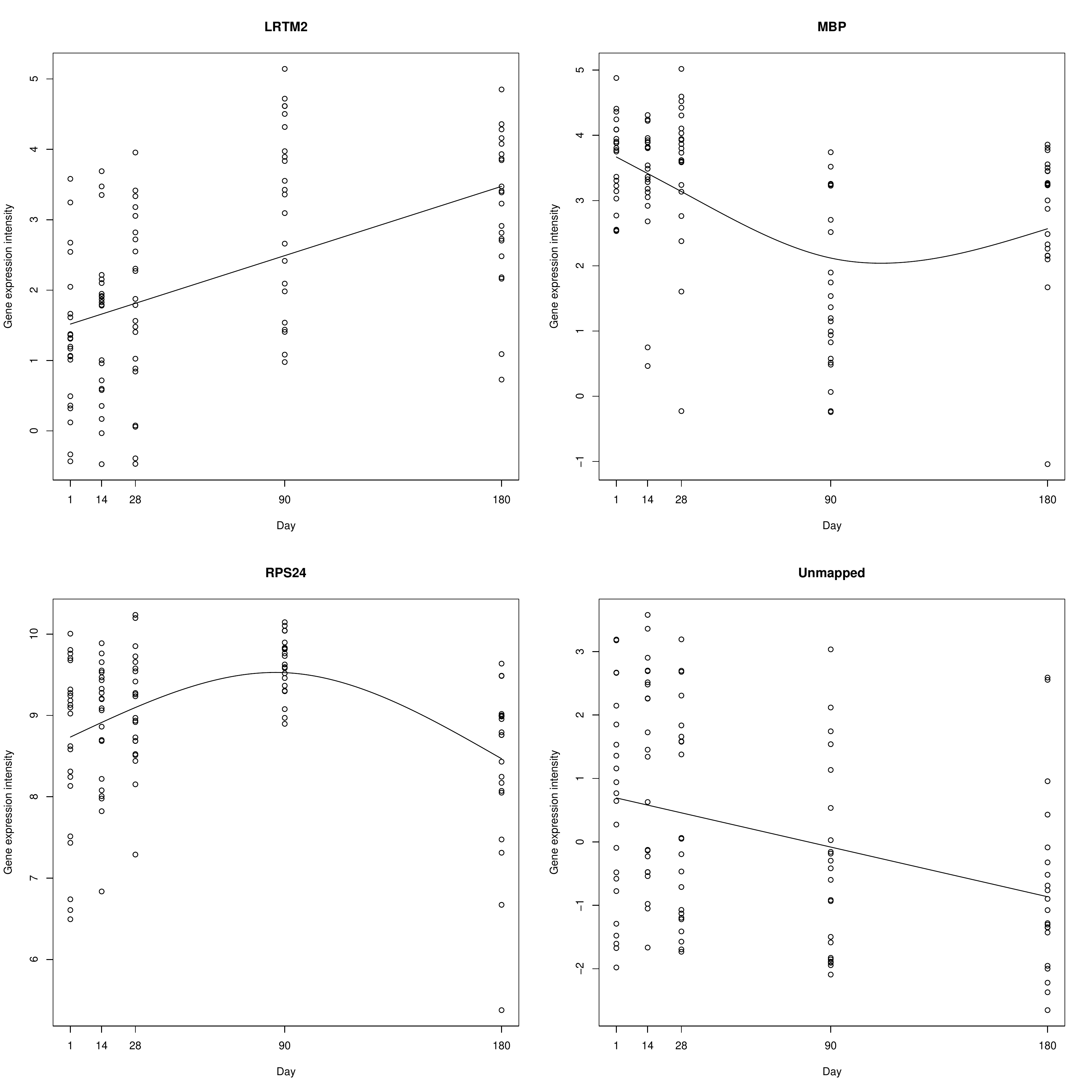}
\caption{\label{fig:outliers}Outlying genes in the fPCA loadings plot shown in Figure \ref{fig:loadings}. These are some of the genes which show the greatest change in expression levels over time}
\end{figure}

\begin{figure}[htbp]
\includegraphics[width=\textwidth]{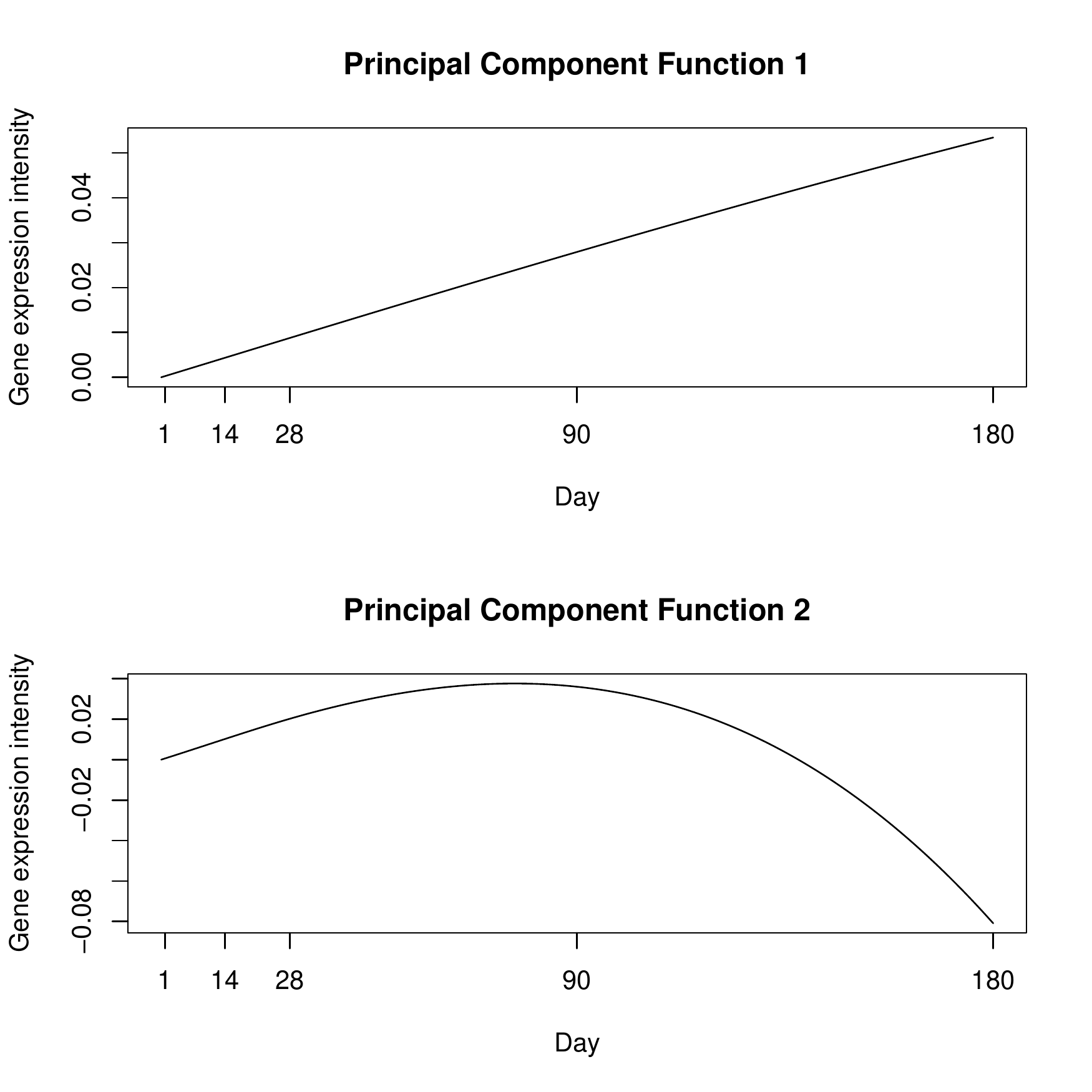}
\caption{\label{fig:functional PCs}Two principal component functions which explain 99.9\% of the variation observed in the fitted mean curves for each gene. The first principal component describes a linear relationship with time. The second principal component captures a more quadratic fit}
\end{figure}

\section{Discussion}\label{sec:discussion}

A number of different models have been proposed in the literature for the analysis of microarray time series data. One of the earliest examples of a FDA approach to the modelling of microarray time series data was \cite{Bar-Joseph2003a} which dealt with the issue of clustering unreplicated data. In their model, the curves were parameterised using B-splines and functional mixed-effects models were used to estimate the cluster mean curves and model the within-cluster variability. In their approach, the function $\mu(\cdot)$ in (\ref{eqn: functional mixed-effects model}) represents a given cluster's mean, and the functions $f_{i}(\cdot), i=1,\cdots,n$ represent the temporal profiles of each of the genes belonging to this cluster, of which there are $n$. A specialised EM algorithm was used to handle dynamic cluster assignments. A very similar approach was developed independently by \cite{Luan2003}.

A problem with \cite{Bar-Joseph2003a,Luan2003} is that the B-spline parameterisation of the curves requires selecting both the number and location of the \textit{knots} - breakpoints for the piecewise polynomials - which control the overall smoothness of the fitted curve $\hat{f}(\cdot)$. As the total number of knots is limited by the number of time points, there is limited scope for controlling the smoothness of the fit. Furthermore, each curve was parameterised using the same number of knots which may be unable to fully capture the wide range of temporal profiles we are likely to observe. \cite{Ma2006} set out to resolve these issues with their alternative framework for clustering. In their model, the cluster mean curves - $\mu(\cdot)$ in (\ref{eqn: functional mixed-effects model}) - are represented using smoothing splines, which place a knot at each design time point and use a roughness penalty to avoid fitted curves which are too `wiggly'. One drawback to their approach, however, is that the individual functions $f_{i}(\cdot), i=1,\cdots,n$ are only modelled as scalar shifts rather than smooth curves. This leads to a more parsimonious model which avoids fitting too many parameters but may fail to adequately model the within-cluster variability. 

\cite{Angelini2009} adopt a fully Bayesian approach to estimation and testing in unreplicated or cross-sectional microarray data sets. Each gene is represented using Legendre polynomials. Three choices for a prior on the noise variance $\sigma^{2}$ allows for errors which are marginally normal, Student \textit{t} or double exponentially distributed, although $\sigma^{2}$ is assumed the same for all genes. This assumption is unlikely to hold in practice, as a correlation between gene expression intensity and measurement noise is well known \citep{Tusher2001}. Given the fully Bayesian framework, hypothesis testing for differences in expression levels across two biological groups is performed using Bayes Factors.

A handful of models and computer packages have also specifically been suggested to model longitudinal data. For instance, \textit{Timecourse} is an R package based on \cite{Tai2006}, where multivariate analysis techniques are applied directly to the vectors of observations. This treatment of time as an unordered categorical variable - found also in ANOVA approaches as in \cite{Wang2003} - has some significant drawbacks. In particular, the method cannot handle missing data, the results obtained by an analysis would be invariant to permutation of the time points, and it is assumed that the time points are regularly spaced. Furthermore, this method only ranks the genes with no guidance given as to how to evaluate significance.

The EDGE method of \cite{Storey2005} is a FDA approach to modelling both longitudinal and cross-sectional microarray data. In their method for longitudinal data analysis, each gene is modelled independently as a separate functional mixed-effects model. The mean curve - $\mu(\cdot)$ in (\ref{eqn: functional mixed-effects model}) - is modelled as a B-spline while the individual effects are treated as scalar shifts as in \cite{Ma2006}. A complete framework for detecting genes differentially expressed across two or more biological groupings is presented, with the model estimation performed by an EM algorithm. Differential expression is quantified using an F-type statistic which compares the residuals of a null model where the biological groupings are ignored to an alternative model where the groupings are taken into account. Significance is assessed by using a resampling bootstrap procedure to estimate the null distribution of this F-type statistic, and the multiple testing problem is handled by analysing the empirical p-value histogram \citep{Storey2003} to estimate the positive false discovery rate.

Another way of accounting for the within-gene variance is to perform a functional principal components analysis (fPCA). This is analagous to the standard PCA, except the principal components (PCs) are functions rather than finite dimensional vectors. There have been a number of different methods suggested for estimating the PCs in a functional context including direct estimation in a mixed-effects model framework \citep{James2000}, standard PCA on discretised curves \citep{Ramsay2005} and `Principal Components Analysis through Conditional Expectation' (PACE) \citep{Yao2005}. It is this latter approach which \cite{Liu2009} applied to the analysis of microarray data; however, PACE was originally proposed for data where the observations on each individual are taken at different time points - for example, in the case of growth curve data - in our experience, microarray experiments tend to have much more regular designs, with each individual observed at the same time points, although these may, indeed, be unequally spaced.

Some key shortcomings of these methods should be noted. Firstly, none of the methods can incorporate the gender and age covariates, particularly as functions of time. Secondly, all of these approachs either use B-splines and/or model the individual `functions' as scalar-shifts, both of which lead to inflexible models. Finally, we are not aware of any existing methods which address the issue of modelling both the within- and between-gene variation.  Our proposed methodology addresses some of these limitations. 

Our results related to the case study presented in section \ref{sec:data} can be compared to the original findings of \cite{Karlovich2009}, who used a non-functional mixed-effect model. \cite{Karlovich2009} lists 19 probes they detected as having a significant gender effect and with a log-transformed signal intensity greater than 7, which we have reproduced here in Table \ref{table:karlovich gender genes}. No justification for this cut-off of 7 is provided, and this filter gives misleading results. For instance, all of the significant gender genes we have identified fail to meet the cut-off. This is because the mean log-transformed signal intensity is taken across both genders, and all of our genes aside from XIST are found on the Y-chromosome and hence completely unexpressed in females.

We were unable to find any confirmation in the literature that TUBB2A is a sex-related gene, and it does not appear in the 15 probes given by \cite{Karlovich2009}. With a mean log-transformed signal intesity of 7.4, it meets their cut-off criteria. It is possible that they removed the probe from their analysis if the residuals from their model were found to be non-normally distributed. Indeed, the unadjusted p-value for the Shapiro-Wilk test on the residuals of \textit{our} model for this probe is $2.54e-5$. However, looking at the residual analysis plotted in Figure \ref{fig:residual analysis}, it is easy to see that there is one very large outlier. This observation corresponds to subject 174 at Day 90. Subject 174 is the individual who developed lung cancer between days 28 and 90, and died prior to day 180. If this observation is removed then the unadjusted Shapiro-Wilk p-value is $0.297$, and the null hypothesis that the residuals are normally distributed is no longer rejected. Hence, TUBB2A may indeed be a novel gender regulated gene.

The number of temporally regulated genes we identified are consistent with \cite{Karlovich2009}, although their method for identifying differentially expressed genes is quite dissimilar to ours (see Section \ref{sec:data}). Indeed, although they found 66 significant genes associated with apoptosis, we found only 15, suggesting the actual significant genes found may vary more widely than the numbers suggest.

\section{Conclusions}
\label{sec:conclusions}

In this paper we have demonstrated a complete framework for the analysis of microarray time series data. The unique characteristics of microarry data lend themselves well to a functional data analysis approach and we have shown how this naturally extends to the inclusion of covariates such as age and sex. Our model presented here is a specialisation of the more general functional mixed-effects model \citep{Rice2001,Guo2002} and, to the best of our knowledge, we are the first to show how to derive the maximum-likelihood estimators, EM-algorithm, confidence intervals and smoother matrix with more than one fixed-effects function.

We were motivated by a real data set and we have aimed to improve upon the existing results with a more flexible model. By taking a roughness penalty approach, this is achieved while avoiding overfitting, allowing for a departure from the original linear mixed-effects model when the data permits it. A deeper biological interpretation is required to fully assess our success here, but the results we have highlighted in this paper suggest that we can easily attach meaning to our findings. It may also prove worthwhile performing a comparative analysis with \cite{Eady2005}, which is another, similar longitudinal study taken over a shorter period of five weeks.

\appendix

\section{Appendix}
\subsection{Example incidence matrix}
\label{appendix:Xi}

In our example data set, there are 5 design time points: Day 1, 14, 28, 90 and 180. Therefore, the incidence matrices for all individuals, $\bm{X}_{i}, i=1,\cdots,n$, all have 5 columns. The first column corresponds to observations at Day 1, the second to observations at Day 14 and so on. The rows correspond to the specific observations on a particular individual. If the individual is observed once at each design time point, then, assuming their vector of observations $\bm{y}_{i}$ has been ordered according to the time points, $\bm{X}_{i} = \bm{I}$.

Now consider the case of subject 174 who died prior to Day 180 and hence only contributed 4 observations at each of the remaining design time points. The design matrix for this individual has 4 rows, corresponding to the 4 observations, but still has 5 columns, corresponding to the design time points. Specifically the incidence matrix in this case is:
\begin{eqnarray*}
\bm{X}_{i} = \left[\begin{array}{ccccc}
1 & 0 & 0 & 0 & 0\\
0 & 1 & 0 & 0 & 0\\
0 & 0 & 1 & 0 & 0\\
0 & 0 & 0 & 1 & 0
\end{array}\right]
\end{eqnarray*}
Note how there is no 1 in the final column which would correspond to an observation at Day 180.

\subsection{Specification of roughness matrix $\bm{G}$}
\label{appendix:G}

\cite{Green1994} show that there is a straight forward way to calculate the roughness matrix for a smoothing spline given the set of distinct time points $\tau_{1}, \cdots \tau_{M}$. The roughness matrix is given as $\bm{G} = \bm{A}\bm{B}^{-1}\bm{A}^{T}$ where the matrices $\bm{A}$ and $\bm{B}$ are defined as follows. First calculate $h_{r} = \tau_{r+1} - \tau_{r}, r=1,\cdots,M-1$, the differences between successive time points. Then matrix $\bm{A}$ is an $M \times (M-2)$ matrix whose entries $a_{r,s}$ are given by
\begin{eqnarray*}
& a_{r,r} = h_{r}^{-1}, \quad a_{r+1,r} = -(h_{r}^{-1} + h_{r+1}^{-1}), \quad a_{r+2,r} = h_{r+1}^{-1} &
\end{eqnarray*}
for $r=1,\cdots,M-2$ and $0$ elsewhere. $\bm{B}$ is an $(M-2) \times (M-2)$ matrix with the entries given by
\begin{eqnarray*}
& b_{1,1} = \frac{h_{1} + h_{2}}{3}, \quad b_{2,1} = \frac{h_{2}}{6} & \\
& b_{r,r+1} = \frac{h_{r+1}}{6}, \quad b_{r+1,r+1} = \frac{h_{r+1} + h_{r+2}}{3}, \quad b_{r+2,r+1} = \frac{h_{r+2}}{6}, \quad r=1,\cdots,M-4 & \\
& b_{M-3,M-2} = \frac{h_{M-2}}{6}, \quad b_{M-2,M-2} = \frac{h_{M-2} + h_{M-1}}{3} &
\end{eqnarray*}

\subsection{Derivation of conditional expectations}
\label{appendix:conditional expectations}

We begin by first considering the posterior expectation of $\bm{\bm{\gamma}_{i}\bm{\gamma}_{i}^{T}}$ which, using
basic properties of expectations, can be rewritten as:
\begin{eqnarray*}
E\left[\frac{1}{n}\sum_{i=1}^{n}\bm{\gamma}_{i}\bm{\gamma}_{i}^{T}|\bm{y},\bm{\eta}=\hat{\bm{\eta}}\right] & = & \frac{1}{n}\sum_{i=1}^{n}E
  \left[\bm{\gamma}_{i}\bm{\gamma}_{i}^{T}|\bm{y},\bm{\eta}=\hat{\bm{\eta}}\right]
\end{eqnarray*}
The definition of covariance allows us to write:
\begin{eqnarray*}
E\left[\bm{\gamma}_{i}\bm{\gamma}_{i}^{T}|\bm{y},\bm{\eta}=\hat{\bm{\eta}}\right] & = &
  E\left[\bm{\gamma}_{i}|\bm{y},\bm{\eta}=\hat{\bm{\eta}}\right]
  E\left[\bm{\gamma}_{i}^{T}|\bm{y},\bm{\eta}=\hat{\bm{\eta}}\right]\\
& + &
  Cov(\bm{\gamma}_{i}|\bm{y},\bm{\eta}=\hat{\bm{\eta}},\bm{\gamma}_{i}^{T}|\bm{y},\bm{\eta}=\hat{\bm{\eta}})
\end{eqnarray*}
The problem is now to determine the mean and covariance of 
 $\bm{\gamma}_{i}|\bm{y}$, for which we use a standard
result conerning the multivariate normal distribution
\citep[See, for example,][]{Anderson1958} which says, for any vectors 
 $\bm{x}_{1}$ and $\bm{x}_{2}$ distributed as
\begin{eqnarray*}
\left[\begin{array}{c}\bm{x}_{1} \\ \bm{x}_{2}\end{array}\right] & \sim &
  \mathcal{N}
  \left(\left[\begin{array}{c}\bm{\mu}_{1} \\ \bm{\mu}_{2}\end{array}\right],
  \left[\begin{array}{cc}\bm{V}_{11} & \bm{V}_{12} \\ 
  \bm{V}_{21} & \bm{V}_{22} \end{array}\right]
  \right)
\end{eqnarray*}
the conditional distribution of $\bm{x}_{1}|\bm{x}_{2}$ is given by
\begin{eqnarray*}
\bm{x}_{1}|\bm{x}_{2} \sim \mathcal{N}[
  \bm{\mu}_{1} + \bm{V}_{12}\bm{V}_{22}^{-1}(\bm{x}_{2} - \bm{\mu}_{2}), 
  \bm{V}_{11} - \bm{V}_{12}\bm{V}_{22}^{-1}\bm{V}_{21}]
\end{eqnarray*}
If we let $\bm{x}_{1} = \bm{\gamma}$ and
  $\bm{x}_{2} = \bm{y}$, and derive the covariance
of $\bm{\gamma}$ and $\bm{y}$ as $Cov(\bm{\gamma},\bm{y}) = \widetilde{\bm{D}}_{\gamma}\widetilde{\bm{X}}^{T}$ then we have
\begin{eqnarray*}
& \left[\begin{array}{c}\bm{\gamma} \\
  \bm{y}|\bm{\eta}=\hat{\bm{\eta}}\end{array}\right] \sim
  \mathcal{N}
  \left(\left[\begin{array}{c}\bm{0} \\ \bm{X}\hat{\bm{\eta}}
  \end{array}\right],
  \left[\begin{array}{cc}\widetilde{\bm{D}}_{\gamma} &  \widetilde{\bm{D}}_{\gamma}\widetilde{\bm{X}}^{T}\\ 
  \widetilde{\bm{X}}\widetilde{\bm{D}}_{\gamma} & \bm{V}\end{array}\right]
  \right)
\\
& \bm{\gamma}|\bm{y},\bm{\eta}=\hat{\bm{\eta}} \sim \mathcal{N}[
  \widetilde{\bm{D}}_{\gamma}\widetilde{\bm{X}}^{T}\bm{V}^{-1}(\bm{y}-\bm{X}\hat{\bm{\eta}}),
  \widetilde{\bm{D}}_{\gamma} -
  \widetilde{\bm{D}}_{\gamma}\widetilde{\bm{X}}^{T}\bm{V}^{-1}\widetilde{\bm{X}}\tilde{\bm{D}}_{\gamma}] &
\end{eqnarray*}
Recognising that, because $\widetilde{\bm{D}}_{\gamma}$ and $\bm{V}$ are block diagonal and 
  $\widetilde{\bm{X}}\widetilde{\bm{D}}_{\gamma}\bm{V}^{-1}(\bm{y}-\bm{X}\hat{\bm{\eta}}) = 
  \hat{\bm{\gamma}}$, we have
\begin{eqnarray*}
\bm{\gamma}_{i}|\bm{y},\bm{\eta}=\hat{\bm{\eta}} \sim \mathcal{N}[
  \hat{\bm{\gamma}}_{i},
  \bm{D}_{\gamma} -
  \bm{D}_{\gamma}\bm{X}_{i}^{T}\bm{V}_{i}^{-1}\bm{X}_{i}\bm{D}_{\gamma}]
\end{eqnarray*}
and we can now write
\begin{eqnarray*}
E\left[\bm{\gamma}_{i}\bm{\gamma}_{i}^{T}|\bm{y},\bm{\eta}=\hat{\bm{\eta}}\right] & = & \hat{\bm{\gamma}}_{i}\hat{\bm{\gamma}}_{i}^{T} +
  [\bm{D}_{\gamma} - \bm{D}_{\gamma}\bm{X}_{i}^{T}\bm{V}_{i}^{-1}\bm{X}_{i}\bm{D}_{\gamma}]
\end{eqnarray*}
For the posterior expectation of $\sigma^{2}$, we follow exactly the same 
approach, writing
\begin{eqnarray*}
& \left[\begin{array}{c}\bm{\epsilon} \\ \bm{y}|\bm{\eta}=\hat{\bm{\eta}}\end{array}\right]
  \sim \mathcal{N}
  \left(\left[\begin{array}{c}\bm{0} \\ \bm{X}\hat{\bm{\eta}}
  \end{array}\right],
  \left[\begin{array}{cc}\bm{R} &  \bm{R}\\ 
  \bm{R} & \bm{V}\end{array}\right]
  \right) &\\
& \bm{\epsilon}|\bm{y},\bm{\eta}=\hat{\bm{\eta}} \sim \mathcal{N}[
  \bm{R}\bm{V}^{-1}(\bm{y}-\bm{X}\hat{\bm{\eta}}),
  \bm{R} - \bm{R}\bm{V}^{-1}\bm{R}] &\\
& \bm{\epsilon}_{i}|\bm{y},\bm{\eta}=\hat{\bm{\eta}} \sim \mathcal{N}[
  \bm{R}_{i}\bm{V}_{i}^{-1}(\bm{y}_{i}-\bm{X}_{i}\hat{\bm{\eta}}),
  \bm{R}_{i} - \bm{R}_{i}\bm{V_{i}}^{-1}\bm{R}_{i}] &
\end{eqnarray*}
Note that
\begin{eqnarray*}
\bm{R}_{i}\bm{V}_{i}^{-1}(\bm{y}_{i}-\bm{X}_{i}\hat{\bm{\eta}}) & = &
  (\bm{V}_{i} - \bm{X}_{i}\bm{D}\bm{X}_{i}^{T})\bm{V}_{i}^{-1}(\bm{y}_{i} -
  \bm{X}_{i}\hat{\bm{\eta}})\\
\nonumber & = & (\bm{I} - \bm{X}_{i}\bm{D}_{\gamma}\bm{X}_{i}^{T}\bm{V}_{i}^{-1})
  (\bm{y}_{i} - \bm{X}_{i}\hat{\bm{\eta}})\\
\nonumber & = & (\bm{y}_{i} - \bm{X}_{i}\hat{\bm{\eta}}) -
  \bm{X}_{i}\bm{D}_{\gamma}\bm{X}_{i}^{T}\bm{V}_{i}^{-1}(\bm{y}_{i} - 
  \bm{X}_{i}\hat{\bm{\eta}})\\
\nonumber & = & \bm{y}_{i} - \bm{X}_{i}\hat{\bm{\eta}} - 
  \bm{X}_{i}\hat{\bm{\gamma}}_{i}\\
\nonumber & = & \hat{\bm{\epsilon}}_{i}\\
\end{eqnarray*}
and
\begin{eqnarray*}
\bm{R}_{i} - \bm{R}_{i}\bm{V_{i}}^{-1}\bm{R}_{i} & = &
  \sigma^{2}\bm{I}_{n_{i}} - \sigma^{4}\bm{V}_{i}^{-1}\\
\nonumber & = & \sigma^{2}(\bm{I}_{n_{i}} - \sigma^{2}\bm{V}_{i}^{-1})
\end{eqnarray*}
and using the identity
\begin{eqnarray*}
E[\bm{\epsilon}_{i}^{T}\bm{\epsilon}_{i}|\bm{y,\bm{\eta}=\hat{\bm{\eta}}}]
  & = & tr\{E[\bm{\epsilon}_{i}\bm{\epsilon}_{i}^{T}|\bm{y},\bm{\eta}=\hat{\bm{\eta}}]\}
\end{eqnarray*}
allows us to derive
\begin{eqnarray*}
E[\bm{\epsilon}_{i}^{T}\bm{\epsilon}_{i}|\bm{y},\bm{\eta}=\hat{\bm{\eta}}]
  & = & tr\{E[\bm{\epsilon}_{i}\bm{\epsilon}_{i}^{T}|\bm{y},\bm{\eta}=\hat{\bm{\eta}}]\}\\
\nonumber  & = & tr\{E[\bm{\epsilon}_{i}|\bm{y},\bm{\eta}=\hat{\bm{\eta}}]E[\bm{\epsilon}_{i}^{T}|\bm{y},\bm{\eta}=\hat{\bm{\eta}}] + \sigma^{2}(\bm{I}_{n_{i}} - \sigma^{2}\bm{V}_{i}^{-1})\}
\\
\nonumber & = & tr\{\hat{\bm{\epsilon}}_{i}\hat{\bm{\epsilon}}_{i}^{T} 
  + \sigma^{2}(\bm{I}_{n_{i}} - \sigma^{2}\bm{V}_{i}^{-1})\}\\
\nonumber & = & tr\{\hat{\bm{\epsilon}}_{i}\hat{\bm{\epsilon}}_{i}^{T}\} +
  tr\{\sigma^{2}(\bm{I}_{n_{i}} - \sigma^{2}\bm{V}_{i}^{-1})\}\\
\nonumber & = & \hat{\bm{\epsilon}}_{i}^{T}\hat{\bm{\epsilon}}_{i} + \sigma^{2}
  (tr\{\bm{I}_{n_{i}}\} - \sigma^{2}tr\{\bm{V}_{i}^{-1}\})\\
\nonumber & = & \hat{\bm{\epsilon}}_{i}^{T}\hat{\bm{\epsilon}}_{i} + \sigma^{2}
  (n_{i} - \sigma^{2}tr\{\bm{V}_{i}^{-1}\})
\end{eqnarray*} 
and so
\begin{eqnarray*}
E[\bm{\epsilon}^{T}\bm{\epsilon}|\bm{y},\bm{\eta}=\hat{\bm{\eta}}] & = & \sum_{i=1}^{n}
  [\hat{\bm{\epsilon}}_{i}^{T}\hat{\bm{\epsilon}}_{i} + \sigma^{2}
  (n_{i} - \sigma^{2}tr\{\bm{V}_{i}^{-1}\})]
\end{eqnarray*}

\bibliography{../../References/references}{}
\bibliographystyle{./Chicago}

\end{document}